\documentclass[pre,twocolumn,showpacs,showkeys,preprintnumbers,floatfix]{revtex4-1}

\usepackage{etex}
\usepackage{ifpdf}
\usepackage{dcolumn}
\usepackage{url}
\usepackage{mathtools}
\usepackage{amsmath}
\usepackage{amscd}
\usepackage{amsfonts}
\usepackage{amssymb}
\usepackage{bm}   
\usepackage{bbm}
\usepackage{verbatim}
\usepackage{stmaryrd}
\usepackage{amsthm}
\usepackage{xcolor}
\usepackage{setspace}
\usepackage{graphicx}
\usepackage{tikz}
\usepackage{braket}
\usetikzlibrary{arrows}
\usetikzlibrary{automata}
\usetikzlibrary{positioning}
\usetikzlibrary{plotmarks}
\usetikzlibrary{patterns}
\usepackage{pgfplots}
\usepgflibrary{patterns}
\pgfplotsset{compat=newest}
\usetikzlibrary{external}
\usepackage{amsmath}
\usepackage{epigraph}

\tikzstyle{vaucanson}=[
  node distance=2cm,
  bend angle=20,
  auto,
  every loop/.style={},
  every edge/.style={->,draw=black,line width=1.2,>=latex,shorten <=1pt,shorten >=1pt},
  every state/.style={draw=black,fill=green!30,line width=2,font=\large},
  loop right/.style={right,out=22,in=-22,loop},
  loop above/.style={above,out=112,in=68,loop},
  loop left/.style={left,out=202,in=158,loop},
  loop below/.style={below,out=292,in=248,loop},
  loop above right/.style={right,out=67,in=23,loop},
  loop above left/.style={left,out=157,in=113,loop},
  loop below left/.style={left,out=247,in=203,loop},
  loop below right/.style={right,out=337,in=293,loop},
]

\tikzset{
  invisiblestate/.style={}
  }

\newcommand{\Symbol}[1]{\textcolor{blue}{#1}}
\newcommand{\Edge}[2]{$#1|\Symbol{#2}$}
\newcommand{\half}{\frac{1}{2}}

\newcommand{\MaxEigBeta}  { {\widehat{\lambda}_\beta} }
\newcommand{\MaxRvecBeta} { {\widehat{\mathbf{r}}_\beta} }

\theoremstyle{plain}    \newtheorem{Lem}{Lemma}
\theoremstyle{plain}    
\theoremstyle{plain}    
\theoremstyle{plain}    
\theoremstyle{plain}    \newtheorem{The}{Theorem}
\theoremstyle{plain}    \newtheorem*{ProThe}{Proof}
\theoremstyle{plain}    
\theoremstyle{plain}    
\theoremstyle{plain}    
\theoremstyle{plain}    
\theoremstyle{plain}    
\theoremstyle{plain}    
\theoremstyle{plain}

\newcommand{\eM}     {\mbox{$\epsilon$-machine}}
\newcommand{\eMs}    {\mbox{$\epsilon$-machines}}
\newcommand{\EM}     {\mbox{$\epsilon$-Machine}}

\newcommand{\Process}{\mathcal{P}}

\newcommand{\MeasAlphabet}  {\mathcal{A}}
\newcommand{\MeasSymbol}   { {X} }
\newcommand{\meassymbol}   { {x} }

\newcommand{\CausalState}   { \mathcal{S} }

\newcommand{\causalstate}   { \sigma }
\newcommand{\CausalStateSet}    { \boldsymbol{\CausalState} }

\newcommand{\Cmu}       {C_\mu}
\newcommand{\hmu}       {h_\mu}

\newcommand{\ProcessAlphabet}   {\MeasAlphabet}

\newcommand{\forward}{+}
\newcommand{\reverse}{-}
\newcommand{\forwardreverse}{\pm} 

\newcommand{\FutureCausalState} { {\CausalState}^{\forward} }

\newcommand{\PastCausalState}   { {\CausalState}^{\reverse} }

\newcommand{\lastindex}[2]{
  \edef\tempa{0}
  \edef\tempb{#2}
  \ifx\tempa\tempb
    \edef\tempc{#1}
  \else
    \edef\tempa{0}
    \edef\tempb{#1}
    \ifx\tempa\tempb
      \edef\tempc{#2}
    \else
      \edef\tempc{#1+#2}
    \fi
  \fi
  \tempc
}

\newcommand{\CSjoint}[1][,]{
   \edef\tempa{:}
   \edef\tempb{#1}
   \ifx\tempa\tempb
      \ensuremath{\FutureCausalState\!#1\PastCausalState}
   \else
      \ensuremath{\FutureCausalState#1\PastCausalState}
   \fi
}

\newif\ifpm
\edef\tempa{\forwardreverse}
\edef\tempb{\pm}
\ifx\tempa\tempb
   \pmfalse
\else
   \pmtrue
\fi

\renewcommand{\H}{\operatorname{H}}

\colorlet {R_color}    {blue}
\colorlet {k_color}    {black!30!green}

\newcommand{\cs}{\causalstate}
\newcommand{\ms}{\meassymbol}
\newcommand{\MS}{\MeasSymbol}

\newcommand{\Abet}{\ProcessAlphabet}

\newcommand{\St}{\CausalState}
\newcommand{\st}{\causalstate}

\def\clap#1{\hbox to 0pt{\hss#1\hss}}

\begin{document}

\title{The Markov Memory for Generating Rare Events}

\author{Cina Aghamohammadi}
\email{caghamohammadi@ucdavis.edu}
\author{James P. Crutchfield}
\email{chaos@ucdavis.edu}

\affiliation{Complexity Sciences Center and Department of Physics, University of
  California at Davis, One Shields Avenue, Davis, CA 95616}

\date{\today}
\bibliographystyle{unsrt}

\begin{abstract}
We classify the rare events of structured, memoryful stochastic processes and
use this to analyze sequential and parallel generators for these events. Given
a stochastic process, we introduce a method to construct a new process whose
typical realizations are a given process' rare events. This leads to an
expression for the minimum memory required to generate rare events. We then
show that the recently discovered classical-quantum ambiguity of simplicity
also occurs when comparing the structure of process fluctuations.
\end{abstract}


\preprint{Santa Fe Institute Working Paper 16-10-XXX}
\preprint{arxiv.org:1610.XXXXX [cond-mat.stat-mech]}

\maketitle 

\setstretch{1.1}


\section{Introduction}

One of the most critical computations today is identifying the statistically
extreme events exhibited by large-scale complex systems. Whether in the domains
of geology, finance, or climate, or whether in natural or designed systems
(earthquakes and hurricanes versus market crashes and internet route flapping),
one can argue that this class of problem is rapidly coming to define our
present scientific and technological era \cite{Crut09b}. Success in
understanding the origins and occurrence of extreme events will have a major
impact on social infrastructure and its sustainability.

Large deviation theory \cite{Deu89,Buck90a,Youn93a,Touc09a,Demb09,Elli12a} is a
relatively new and key tool for analyzing a process' full range of statistical
fluctuations---in particular, those well outside the domain of the Law of Large
Numbers. Presaged by Shannon-McMillman-Breiman type theory in communication
theory \cite{Cove06a,Algo89}, the mathematical development of large deviations
was first pursued by Donsker and Varadhan \cite{Donsk75}. In essence, it can be
seen as a refinement of the Central Limit Theorem \cite{Fell70a} or as a
generalization of Einstein's fluctuation theory \cite{Eins05a,Eins06a}. Today,
large deviation theory enters into physics in many different circumstances
\cite{Elli12a}. One can also formulate statistical mechanics in the language of
large deviation theory \cite{Oono89a}. And it appears in abstract dynamical
systems under the rubric of the thermodynamic formalism \cite{Ruel78}.

The following analyzes the memory resources required to generate, and so study,
extreme events in structured temporal processes. It extends large deviation
theory in a constructive way that leads to exact calculations of the spectrum
of fluctuations for processes generated by finite-state hidden Markov models.
Fortunately, in this setting the generation and fluctuation problems can be
simply stated. And so, we first give a suitably informal introduction to
process generators and fluctuation theory, leaving technical results for later.

\section{Optimal serial and parallel generators}

To keep matters uncomplicated, consider a process consisting of time series
$\ldots 10010011 \ldots$ of binary symbols. Having raw sequences in hand does
represent the process' behaviors, but in and of themselves the sequences are
not that useful. For example, how can we predict future symbols? What
mechanisms drive the process' behaviors? Much more helpful in answering such
questions is a model that can produce the process' sequences. And, a good one
can be used to simulate the process---generating example sequences, perhaps not
even in the original data, but statistically similar---that allow one to
predict future sequences, gain insight into the process' internal mechanisms,
and estimate statistical properties.

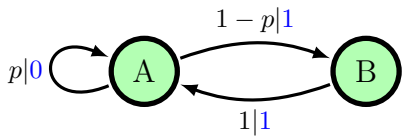
\begin{figure}
  \centering
  \begin{tikzpicture}[style=vaucanson]
    \node [state] (A)                   {A};
    \node [state] (B)  [right=2cm of A] {B};

    \path (A) edge [loop left]   node        {~\Edge{p}{0}}   (A)
          (A) edge [bend left]  node { \Edge{1-p}{1}} (B)
          (B) edge [bend left]  node { \Edge{1}{1} }  (A);
  \end{tikzpicture}
\caption{State-transition diagram for the hidden Markov generator of the Even
	Process, which consists of random binary sequences with an even number of
	$1$s separated by arbitrary-length blocks of $0$s.
	}
\label{fig:HMM_EP}
\end{figure}

Markov chains (MCs) \cite{Norr98,LEVIN09} and hidden Markov models (HMMs)
\cite{Rabi86a,Rabi89a,Uppe97a} are widely used to generate stochastic
processes. Both kinds of model consist of a set $\CausalStateSet$ of states and
a set of state transitions. With HMMs process symbols are distinct from the
internal states, whereas in MCs they are synonymous. Figure~\ref{fig:HMM_EP}
gives an example: the state-transition diagram for a two-state HMM that
generates the binary-symbol \emph{Even Process} \cite{Crut01a}. The Even
Process highlights why HMMs are such useful representations. Since the process
symbols are not the states, HMMs can be arbitrarily more compact than MCs for
the same process. In this case, the Even Process is an infinite Markov order
process since its current state can depend on arbitrarily long histories. (If
only $1$s have been observed, it can be in either state $A$ or state $B$.) Said
in terms of model size, the MC representing the Even Process requires an
infinite number of Markov states, each associated with a history $1^k 0$, $k =
0, 1, 2, \ldots$. In contrast, as the figure shows, the Even Process' HMM takes
only two states.

When using HMMs as process generators we can restrict attention to those that
are \emph{unifilar}: the current state and next symbol uniquely determine the
next state. Unifilar HMMs are important since they are perfect predictors of
their process. (The same is not generally true of a process' nonunifilar HMM
generators. We return to the important, but subtle distinction between
prediction and generation using HMMs at the end.) For any given process there
is an infinite number of unifilar HMM generators; so the restriction imposes no
loss of representational generality.  Given all of the alternative HMMs,
though, which do we choose?

Let's say Alice wants to generate the Even Process. To do this, she writes a
computer program: If the current state is $A$, with probability $p$ the program
emits symbol $0$ and stays at state $A$ and with probability $1-p$ it emits
symbol $1$ and goes to state $B$. However, if the current state is $B$, it
generates symbol $1$ and goes to the state $A$. The program continues in this
fashion, again and again, and in the long run generates a realization of the
Even Process. Moreover, if Alice chooses to start in $A$ or $B$ using the
asymptotic state probability distribution $\pi$, then the resulting realization
is stationary.

Imagine that a long time has passed and the HMM is in state $A$. Alice decides
to stop the program for now and return tomorrow to continue generating the same
realization. Now, she must make a decision, does she use the realization
generated today or start all over again tomorrow? Not wanting to waste the
effort already invested, she decides to use today's realization tomorrow and
simply concatenate newly generated symbols.

The next day, though, can she randomly pick a state and continue generating?
The answer is no. If she randomly picks state $B$, then there is a chance that
after concatenating the old and new realizations together, the sequence has odd
number of $1$s between two $0$s. However, she knows that the Even Process never
generates such subsequences. Thus, if she wants to use today's realization
tomorrow then, she must record the HMM's current state and continue generating
from that state tomorrow \footnote{The time period over which Alice pauses
generation can be set to any duration---an hour, a minute, or a second. In
particular, the period can be that required to generate a single symbol. In
this case, after every symbol emitted Alice must know in what state the
generator is. In short, Alice needs to remember the current state during
generation.}.

Information theory \cite{Cove06a} tells us that to record the current state
Alice needs $\log_2 |\CausalStateSet|$ bits of memory. This is the cost of
\emph{sequential generation}. And, it gives a quantitative way to compare
models across the infinite number of alternatives. If Alice wants to use less
memory, she selects the HMM with the minimum number of states.  Which
representation achieves this?

Before answering, let's contrast another scenario, that for \emph{simultaneous
generation}.  Now, Alice wants to generate $N$ realizations for a given process
simultaneously, but insists that the individual sequences to be statistically
independent. The latter means that she cannot simply generate a single
realization and copy it $N$ times. At first blush, it seems that she needs $N
\log_2 |\CausalStateSet|$ bits of memory. According to Shannon's source coding
theorem \cite{Shan48a,Cove06a}, though, she can compress the sequence
information and, for large $N$, she needs only $N \H[\CausalStateSet] \leq N
\log_2 |\CausalStateSet|$ bits of memory, where $\H[\CausalStateSet] =
-\sum_{\cs \in \CausalStateSet} \pi(\cs) \log_2 \pi(\cs)$ is the Shannon
entropy of the stationary probability distribution $\pi(.)$ over the HMM's
states. That is, on average Alice needs $\H[\CausalStateSet]$ bits of memory to
generate each realization. So, if Alice wants to use less memory, she selects
the process HMM with the minimum $\H[\CausalStateSet]$ in the set of unifilar
HMMs. Again, which representation achieves this?

Crutchfield and Young \cite{Crut88a} showed that over all unifilar HMMs that
generate a given process, there is unique HMM with the minimum number of
states. Surprisingly this same HMM is also the one with the minimum entropy
over it's states. It is now known as the \emph{\eM} \cite{Crut12a,Shal98a} and
it's state entropy is the process' \emph{statistical complexity} $\Cmu$
\cite{Crut88a,Crut12a}. The consequence is that, for a given stochastic
process, the minimum memory required for any unifilar HMM to sequentially
generate it is $\log_2 |\CausalStateSet_{\epsilon}|$ bits, where
$\CausalStateSet_{\epsilon}$ is the set of states in the process' \eM. And, for
simultaneous generation the average minimum required memory for each
realization is $\Cmu$.

Today, $\Cmu$ is often used as a measure of structural complexity for
stochastic processes, from stochastic resonance \cite{Witt97a} to hydrodynamic
flows \cite{Gonc98a}, atmospheric turbulence \cite{Palm00a}, geomagnetic
volatility \cite{Clar03a}, and single-molecule dynamics
\cite{Li13a,Nerukh10,Kell12a}. In short, we use \eMs\ and $\Cmu$ to measure the
memory inherent in a stochastic process. And, by the preceding argument we now
know how they determine the memory required for sequential and parallel
generation.

\section{Typical and Atypical Behaviors}

So far, the discussion implicitly assumed that models captured a process'
typically observed behaviors. However, most stochastic processes exhibit
statistical fluctuations and so occasionally generate atypical, statistically
extreme behaviors. Now, we turn to define what we mean by typical and atypical
behaviors. Once done, we finally state our problem: How much memory is needed
to generate a process' atypical behaviors.

We need to backtrack a bit to define a process more carefully. A discrete-time,
discrete-value \emph{stochastic process} \cite{Travers13,Uppe97a} is the
probability space $\Process=\big\{\MeasAlphabet^\infty, \Sigma, \mathbb{P}(\cdot)
\big\}$. Here, $\mathbb{P}(\cdot)$ is the probability measure over the
bi-infinite chain $\MS_{-\infty:\infty} = \ldots \MS_{-2} \MS_{-1} \MS_0 \MS_1
\MS_2 \ldots$, where random variables $\MS_i$ take values in a finite discrete
alphabet $\MeasAlphabet$ and $\Sigma$ is the $\sigma$-algebra generated by the
cylinder sets in $\MeasAlphabet^\infty$. The following only considers ergodic
stationary processes; that is, $\mathbb{P}(\cdot)$ is invariant under time
translation---$\mathbb{P}(\MS_{i_1}\MS_{i_2} \cdots \MS_{i_m}) =
\mathbb{P}(\MS_{i_1+n}\MS_{i_2+n} \cdots \MS_{i_m+n})$ for all $n$---and over successive
realizations.

So, what does it mean that a process exhibits statistical fluctuations? Let's
say Alice has a biased coin, meaning that when she flips it, the probability
$p$ of seeing heads is greater than one half. Alice now flips the coin $n \gg
1$ times and see $k$ heads. The Strong Law of Large Numbers \cite{Durr10}
guarantees that for large $n$, the ratio $k/n$ almost surely converges to $p$:
\begin{align*}
\mathbb{P} \left( \lim_{n \to \infty} \frac{k}{n} = p \right) = 1
  ~. 
\end{align*}
Informally, for large $n$ the \emph{typical sequence} has close to $p$ percent
Heads. This does not mean that Alice never sees long runs of all Heads or all
Tails, for example. It simply means that the latter are rare events.

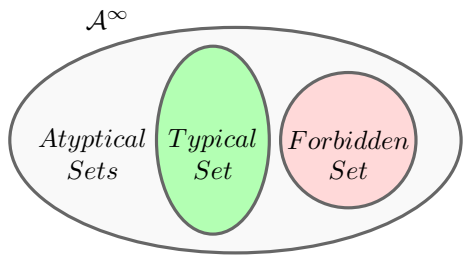
\begin{figure}
\begin{tikzpicture}
\def\R{4}
\filldraw[color=black!60, fill=gray!5, very thick](2.5,0) ellipse (3 and 1.5);
\filldraw[color=black!60, fill=green!30, very thick](2.2,0) ellipse (0.75 and 1.25);
\filldraw[color=black!60, fill=red!15, very thick](4.0,0) ellipse (0.9 and 0.9);
\node at (2.2,0) {$Typical$};
\node at (2.2,-0.4) {$Set$};
\node at (0.6,0) {$Atyptical$};
\node at (0.6,-0.4) {$Sets$};
\node at (4.0,0) {$Forbidden$};
\node at (4.0,-0.4) {$Set$};
\node at (0.8,1.6) {$\Abet^\infty$};
\end{tikzpicture}
\caption{For a given process, the space $\Abet^\infty$ of its realizations is
	partitioned into forbidden sequences, sequences in the typical set, and
	sequences in atypical sets.
  }
\label{fig:TSNTS}
\end{figure}

We now show that a process' typically observed realizations are those sequences
in its so-called typical set. Consider a given process and let
$\MeasAlphabet^n$ denote the set of length-$n$ sequences. Then, for an
arbitrary $\epsilon > 0$ the process' \emph{typical set}
\cite{Cove06a,Kull68,Yeun08a} is:
\begin{align}
A_{\epsilon}^{n} \! = \! \{ w:
  2^{-n (\hmu + \epsilon)} \leq \mathbb{P}(w)
  \leq 2^{-n (\hmu - \epsilon)}, w \in \MeasAlphabet^n \}
  ,
\label{eq:TSDEF}
\end{align}
where $\hmu$ is the process' \emph{metric entropy} (Shannon entropy rate)
\cite{Han06}:
\begin{align*}
\hmu(\Process) = - \lim_{n \to \infty} \frac{1}{n} \sum_{w \in \Abet^n } \mathbb{P}(w) \log_2 \mathbb{P}(w)
  ~.
\end{align*}
According to the \emph{Shannon-McMillan-Breiman theorem}
\cite{Shan48a,McMi53a,Brei57}, for a given $\epsilon \ll 1$ and sufficiently large $n$:
\begin{align}
\mathbb{P}(w \notin A_{\epsilon}^{n}, w \in \MeasAlphabet^n) \leq \epsilon
  ~.
\label{eq:AtypProb}
\end{align}
There are two important lessons here. First, coming from Eq. (\ref{eq:TSDEF}),
all sequences in the typical set have approximately the same probability.
Second, coming from Eq. (\ref{eq:AtypProb}), for large $n$ the probability of
sequences falling outside the typical set is close to zero---they are rare.

One consequence is that sequences generated by a stationary ergodic process
fall into one of three partitions; see Fig.~\ref{fig:TSNTS}. The first contains
those that are never generated by a process---sequences with zero probability.
(For example, the Even Process cannot generate realizations containing a subsequence
in $\{01^{2k+1}0\}$, $k = 0, 1, 2, \ldots$---those with an odd number of $1$s
between $0$s.) These are the \emph{forbidden sequences}. The second partition
consists of those in the typical set---the set with probability close to one,
as in Eq. (\ref{eq:TSDEF}). And, the last contains sequences in a family of
atypical sets---realizations that are rare to different degrees. We now refine
this classification.

\begin{figure}
\centering
\includegraphics[width=0.8\columnwidth]{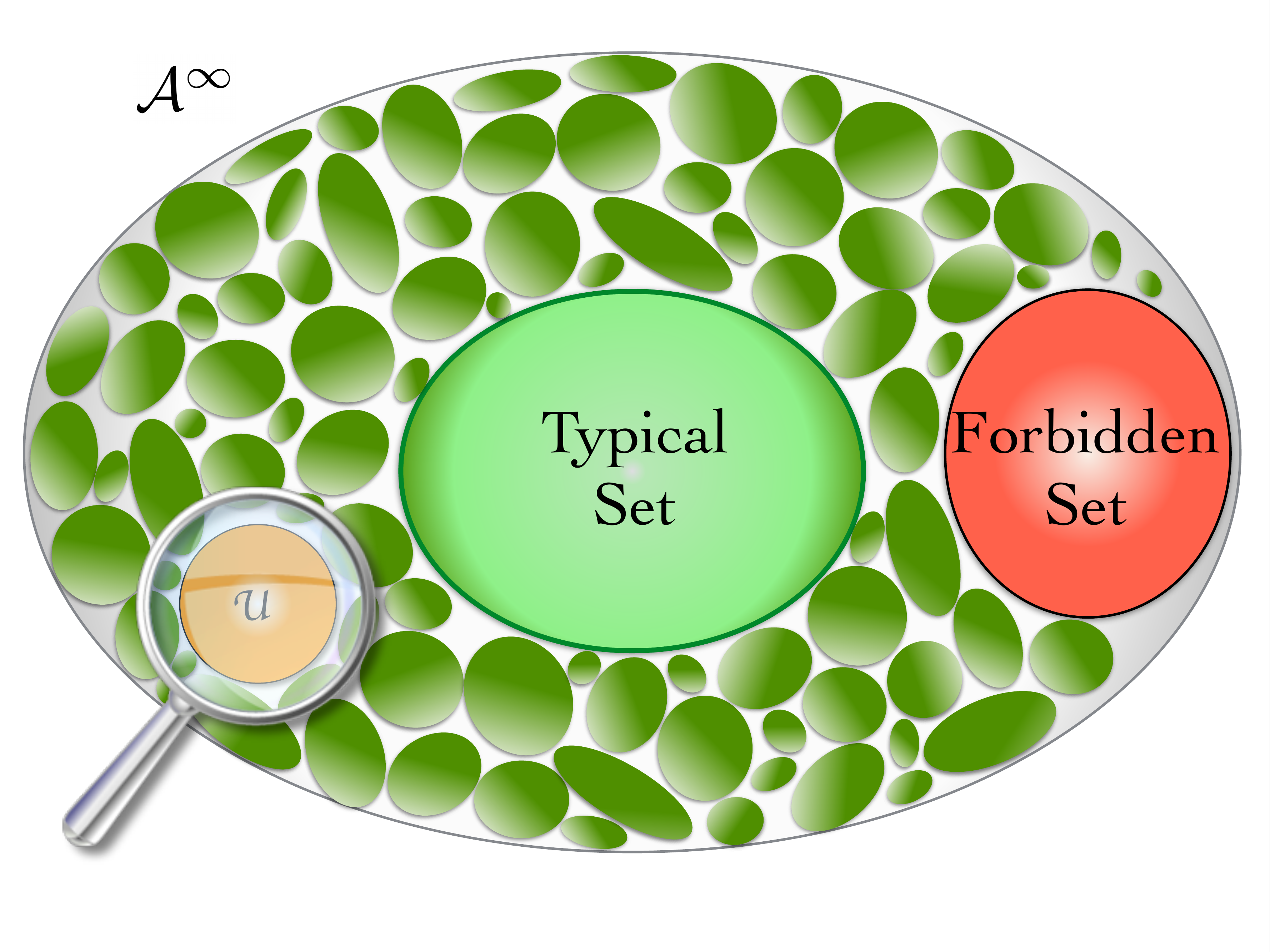}
\caption{$\MeasAlphabet^\infty$ partitioned into $\Lambda_U$s---isoenergy
	or equal probability-decay-rate bubbles---in which all sequences in the same
	$\Lambda_U$ have the same energy $U$. The typical set is one such bubble
	with energy equal to metric entropy: $U = \hmu$. Another important
	partition is that of the forbidden sequences, in which all sequences have
	zero probability. The forbidden set can also be interpreted as the subset of
	sequences with infinite energy. 
	}
\label{BUBEN}
\end{figure}

Mirroring the familiar \emph{Boltzmann weight} in statistical physics
\cite{BOLT12}, in the $n \to \infty$ limit, we define the subsets $\Lambda^\Process_U
\subset \MeasAlphabet^\infty$ for a process $\Process$ as:
\begin{align}
\Lambda^\Process_{U,n} & = \left\{w: -\frac{\log_2\mathbb{P}(w)}{n} = U ,
	\ w \in \MeasAlphabet^n \right\}
	\nonumber \\
    \Lambda^\Process_U &= \lim_{n \to \infty} \Lambda_{U,n}
  ~.
\label{ENERGY}
\end{align}
In effect, this partitions $\MeasAlphabet^\infty$ into subsets
$\Lambda^\Process_U$ in which all $w \in \Lambda^\Process_U$ have the same
probability decay rate $U$. Physics vernacular would speak of the sequences
having the same \emph{energy density} $U$. Figure~\ref{BUBEN} depicts these
subsets as ``bubbles'' of equal energy. (Though, to be clear about their
``shape'', these subsets are isomorphic to Cantor sets.) The definition
guarantees that any bi-infinite sequence $\Process$ generates belongs to one of
these sets. Equation~(\ref{eq:TSDEF}) says the typical set is that bubble with
energy equal to the process' entropy rate: $U = \hmu$. All the other bubbles
contain rare events.

When Alice uses a process' HMM to generate realizations, what she does is
generate sequences in the typical set with probability close to one and,
rarely, atypical sequences. Imagine, though, that Alice is interested in a
particular class of rare sequences, those in a different isoenergy bubble; say,
those with energy $U$ in the set $\Lambda^\Process_U$. How can Alice
efficiently generate these rare sequences? We now show that she can find a new
process $\Process^U$ whose typical set is $\Lambda^\Process_U$.

\section{Generating rare events}

To do this, we return to considering models for a given process. With suitable
models and a precise definition of a process' atypical sequences we can now
ask, How much memory is required to generate them? How does this compare to the
memory required to generate typical behaviors? Before providing the answers,
let's first revisit HHMs and remotivate using them.

Note that in most cases representing a process by specifying the probability
measure $\mathbb{P}(\cdot)$ is impossible due to the infinite number of
possible sequences. So, how should we represent processes? Is there a more
compact way than specifying in-full the probability measure on the sequence
sigma algebra? In a rather direct sense, Markov chains and hidden Markov models
provide constructive answers. The quality of those answers depends, of course,
on how useful these representations are. We now fill in their technical
details, so that we can work with them.

A \emph{hidden Markov model} (HMM) is a tuple $\big\{ \CausalStateSet,
\MeasAlphabet, \{ T^{(\ms)}, \ms \in \MeasAlphabet \} \big\}$. In this,
$\CausalStateSet$ is a finite set of states, $\MeasAlphabet$ is a finite
alphabet, and $\{T^{(\ms)}, \ms \in \MeasAlphabet\}$ is a set of
$|\CausalStateSet| \times |\CausalStateSet|$ substochastic symbol-labeled
transition matrices whose sum $T= \sum_{\ms \in \MeasAlphabet} T^{(\ms)}$ is an
stochastic matrix. Consider an example HMM where $\CausalStateSet=\{A,B\}$,
$\MeasAlphabet= \{ 0,1\}$, 
$T^{(0)} = 
 \begin{bmatrix}
    p & 0 \\
    0 & 0
\end{bmatrix}$,
and 
$T^{(1)} = 
 \begin{bmatrix}
    0 & 1-p \\
    1 & 0
\end{bmatrix}$.
An HMM such as this is graphically depicted via its state-transition
diagram---a directed graph with labeled edges. $\CausalStateSet$ is the set of
graph nodes and the edge from node $i$ to $j$ is labeled by $p | \ms$
corresponding to the HMM transition with probability $p = T^{(\ms)}_{ij}$ that
goes from state $i$ to $j$ and generates symbol $\ms$. This HMM, in fact, is
the example we have been using: the Even Process HMM shown in
Fig.~\ref{fig:HMM_EP}. Since it is minimal and unifilar it is the process' \eM.

Given a process $\Process$ and its \eM\ $M(\Process)$, what is the \eM\
$M(\Process^U)$ that generates the atypical sequences $\Process^U$ at some
energy $U \neq \hmu$? Here, we answer this question by constructing a map
$\mathcal{B}_\beta: \Process \rightarrow \Process_\beta$ from the original
$\Process$ and a new one $\Process_\beta$ for which we introduce a new
parameter $\beta \in \mathbb{R} /\{0\}$ that indexes the atypical set of
interest. Both processes $\Process =\big\{\MeasAlphabet^\infty,
\Sigma,\mathbb{P}(\cdot) \big\}$ and
$\Process_\beta=\big\{\MeasAlphabet^\infty, \Sigma,\mathbb{P}_\beta(\cdot)
\big\}$ are defined on the same measurable sequence space. The measures differ,
but their supports (allowed sequences) are the same. We refer to
$\mathcal{B}_\beta$ as the \emph{$\beta$-map}.

Assume we are given $M(\Process) = \big\{ \CausalStateSet,  \MeasAlphabet, \{
T^{(\ms)}, \ms \in \MeasAlphabet \} \big\}$. We will now show that for every
probability decay rate or energy $U$, there exists a particular $\beta$ such
that $\Process^U = \Process_\beta$. The $\beta$-map which establishes this is
calculated by a construction that relates $M(\Process)$ to $M(\Process_\beta)
=\big\{ \CausalStateSet, \MeasAlphabet, \{ \mathbf{S}_\beta^{(\ms)}, \ms \in
\MeasAlphabet \} \big\}$---the HMM that generates $\Process_\beta$:
\begin{enumerate}
\setlength{\topsep}{-5pt}
\setlength{\itemsep}{-5pt}
\setlength{\parsep}{-5pt}
\item For each $\ms \in \MeasAlphabet$, construct a new matrix
	$\mathbf{T}^{(\ms)}_\beta$ for which $\big( \mathbf{T}^{(\ms)}_\beta
	\big)_{ij} = \big( \mathbf{T}^{(\ms)} \big)_{ij}^\beta$.
\item Construct a new matrix  $\mathbf{T}_\beta = \sum_{\ms \in \Abet}
	T^{(\ms)}_{\beta}$.
\item Calculate $\mathbf{T}_{\beta}$'s maximum eigenvalue $\MaxEigBeta$ and
	corresponding right eigenvector $\MaxRvecBeta$.
\item For each $\ms \in \MeasAlphabet$, construct new matrices
	$\mathbf{S}_\beta^{(\ms)}$ for which:
  \begin{align}
	 \big( {\textbf S}^{(\ms)}_\beta \big)_{ij}
  	= \frac {\big(\mathbf{T}^{(\ms)}_\beta \big)_{ij} (\MaxRvecBeta)_j }
  	{\MaxEigBeta (\MaxRvecBeta)_i }
  	~.
  \label{eq:TiltedS}
  \end{align}
\end{enumerate}

{\The \label{THEO} $\Process_\beta = \Process^U$, where:
\begin{align}
U = \beta^{-1} \big(\hmu(\Process_\beta) - \log_2 \MaxEigBeta \big)
  ~.
\end{align}
}

{\ProThe See the appendix.}

This says that changing $\beta$ controls which class of rare events we focus
on. Informally, the $\beta$-map acts like a magnifier (Fig.~\ref{BUBEN}) by
enhancing particular isoenergy bubbles. That is, changing $\beta$ moves the
magnifier from one bubble to another. The $\beta$-map construction guarantees
that the HMMs $M(\Process)$ and $M(\Process_\beta)$ have the same states and
transition topology: $\big( \mathbf{T}^{(\ms)}_\beta \big)_{ij} \neq 0  \iff
\big( {\textbf S}^{(\ms)}_\beta \big)_{ij} \neq 0$. The only difference is in
their transition probabilities. Thus, $M(\Process_\beta)$ is also a unifilar
HMM, but not necessarily an \eM, since the latter requires a minimal set of
states. Minimality is not guaranteed by the $\beta$-map. Typically, though,
$M(\Process_\beta)$ is an \eM\ and there is only a finite number of $\beta$s for
which it is not. (More detailed development along these lines will appear in a
sequel.)

\begin{figure}
  \centering
  \begin{tikzpicture}[style=vaucanson]
    \node [state] (A)       [fill=yellow!40]            {A};
    \node [state] (B)  [below=2cm of A, fill=yellow!40] {B};
    \node [state]      (D)   [right=3cm of A, fill=yellow!40]                               {D};
    \node [state]      (E)   [ below left=2.3cm and 1cm of D, fill=yellow!40]  {E};
    \node [state]      (F)   [below right=2.3cm and 1cm of D, fill=yellow!40] {F};
\draw[black,thick] (1.6,-3.5) -- (1.6,0.5);
    \path (A) edge [loop left]   node        {~\Edge{p}{0}}   (A)
          (A) edge [bend left]  node { \Edge{1-p}{1}} (B)
          (B) edge [loop left]  node { \Edge{p}{1} }  (B)
          (B) edge [bend left]  node { \Edge{1-p}{0} }  (A)
          (D)   edge [loop below] node        {~\Edge{$p$}{1}} (D)
          (D)   edge [bend right]       node [swap] { \Edge{$1-p$}{0}} (E)
          (E)   edge [bend right]       node [swap] { \Symbol{1}} (F)
          (F)   edge [bend right]       node [swap] { \Symbol{1}} (D);
  \end{tikzpicture}
\includegraphics[width=1.1\columnwidth]{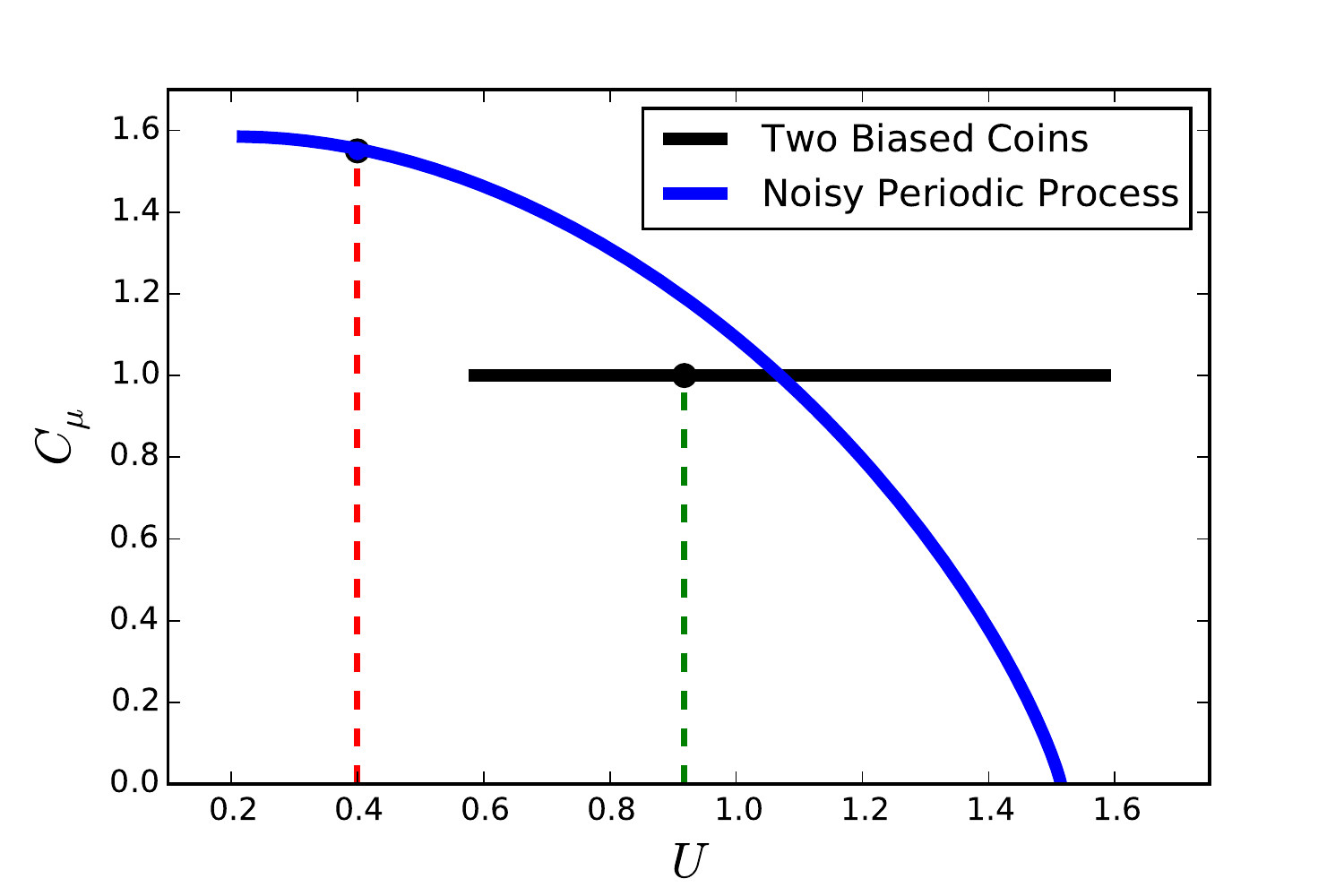}
\caption{Upper left: \EM\ generator of the Two-Biased Coins Process. Upper
	right: \EM\ generator of the Intermittent Periodic Process. Bottom: Statistical
	complexity $\Cmu$ versus energy $U$ (or fluctuation class) for each, along
	with the energies $U^*$ at which their typical sets are found (vertical
	dashed lines).
	}
\label{HMMPC}
\end{figure}
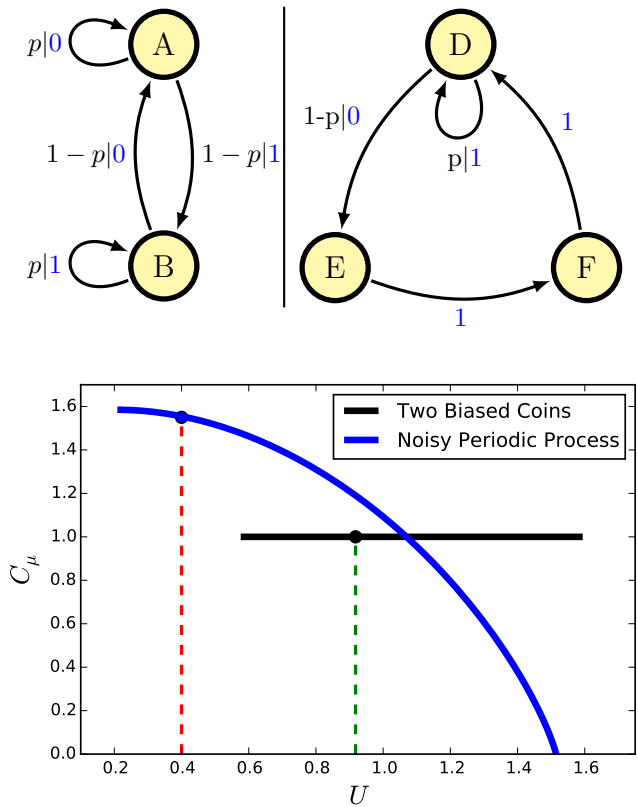

\section{Memory Spectra}

For an arbitrary stochastic process $\Process$, the last section presented a
method to construct a (unifilar) generator whose typical set is the process
$\Process^U$---the rare events of the original $\Process$. Now, we determine
the minimum memory required to generate $\Process^U$. Recalling the earlier
coding-theoretic arguments, this is rather straightforward to answer. The
minimum memory to generate $\Process^U$ is determined by the size of its \eM.
(As noted, this is the size of $M(\Process_U)$ except for finite number of $U$.)

And so, except for a finite number of rare-event classes, to sequentially
generate sequences in a given rare class, one requires the same memory---the
number $|\CausalStateSet|$ of states---as that to generate the original process.
This is our first result on required Markov memory for a process' rare events.

The story differs markedly, however, for simultaneous generation. The minimum
required memory for simultaneous generation of $\Process^U$ is
$\Cmu(\Process^U)$, putting the earlier coding argument together with last
section's calculations. More to the point, this is generally not equal to
$\Cmu(\Process)$. To better appreciate this result, let us examine three
examples.

First, consider the Two-Biased Coins (TBC) Process with $p= 1 / 3$, whose \eM\
is shown in Fig.~\ref{HMMPC}(top left). To generate its realizations one flips
a biased coin repeatedly. At first, label Heads a $0$ and Tails a $1$. After
flipping, switch the labels and call a Head $1$ and Tail $0$. A TBC process
sequence comes from repeating these steps endlessly.  As Fig.~\ref{HMMPC} makes
plain, there is a symmetry in the process. In the stationary distribution
$\pi$, state $A$ has probability half, as does state $B$, and this is
independent of $p$. This gives $\Cmu(\Process) = 1$ bit. Recalling the
$\beta$-map construction, we see that changing $\beta$ does not change the \eM\
topology. All that changes is $p$. This means, in turn, that the symmetry in
states remains and $\Cmu(\Process^U) = 1$ is constant over allowed $U$s (or
$\beta$s); $\Cmu(U)$ versus $U$ is the horizontal line shown in
Fig.~\ref{HMMPC}.

What energies are allowed? The TBC Process has a finite energy range: $U \in
[\approx 0.586,\approx 1.584]$. From
Eq.~(\ref{ENERGY}) we see that the maximum $U_\text{max}$ corresponds to the
bubble with the rarest sequences that can be generated. Conversely, 
$U_\text{min}$ corresponds to the bubble with the most probable. The
energy extremes delimit the domain of the $\Cmu(\Process^U)$ curves in Fig.
\ref{HMMPC}. In addition, the $U$ associated with $\Process$'s typical set is
marked in the figure with a dashed (green) vertical line near $U \approx 0.9183$.

The difference between the typical set and that with $U_\text{min}$ is
important to appreciate. The typical set is that \emph{set} of sequences with
probability close to one and with energy $U = \hmu$. The latter is generally
different from $U_\text{min}$. That is, typical sequences are not necessarily
the most probable sequences, considered individually, but rather they belong
to the most probable subset---the typical set.

As a result of this analysis, for this example, independent of which class of
rare events we examine, one $1$ bit of memory is uniformly required for
generating the TBC Process' events, rare or not.

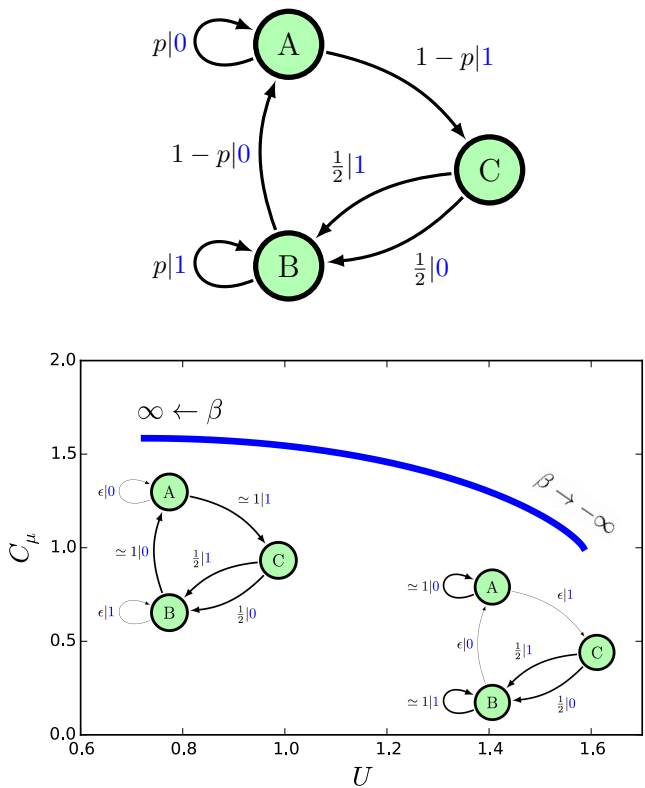
\begin{figure}
  \centering
  \begin{tikzpicture}[style=vaucanson]
    \node [state] (A)                   {A};
    \node [state] (B)  [below=2cm of A] {B};
    \node [state]      (C)   [below right=1 cm and 2cm of A]   {C};
    \path   (A) edge [bend left]  node { \Edge{1-p}{1}} (C)
          (A) edge [loop left]   node        {~\Edge{p}{0}}   (A)
          (B) edge [loop left]  node { \Edge{p}{1} }  (B)
          (B) edge [bend left]  node { \Edge{1-p}{0} }  (A)
          (C) edge [bend right]  node [swap] { \Edge{\half}{1} }  (B)
          (C) edge [bend left]  node { \Edge{\half}{0} }  (B);
  \end{tikzpicture}
\includegraphics[width=1\columnwidth]{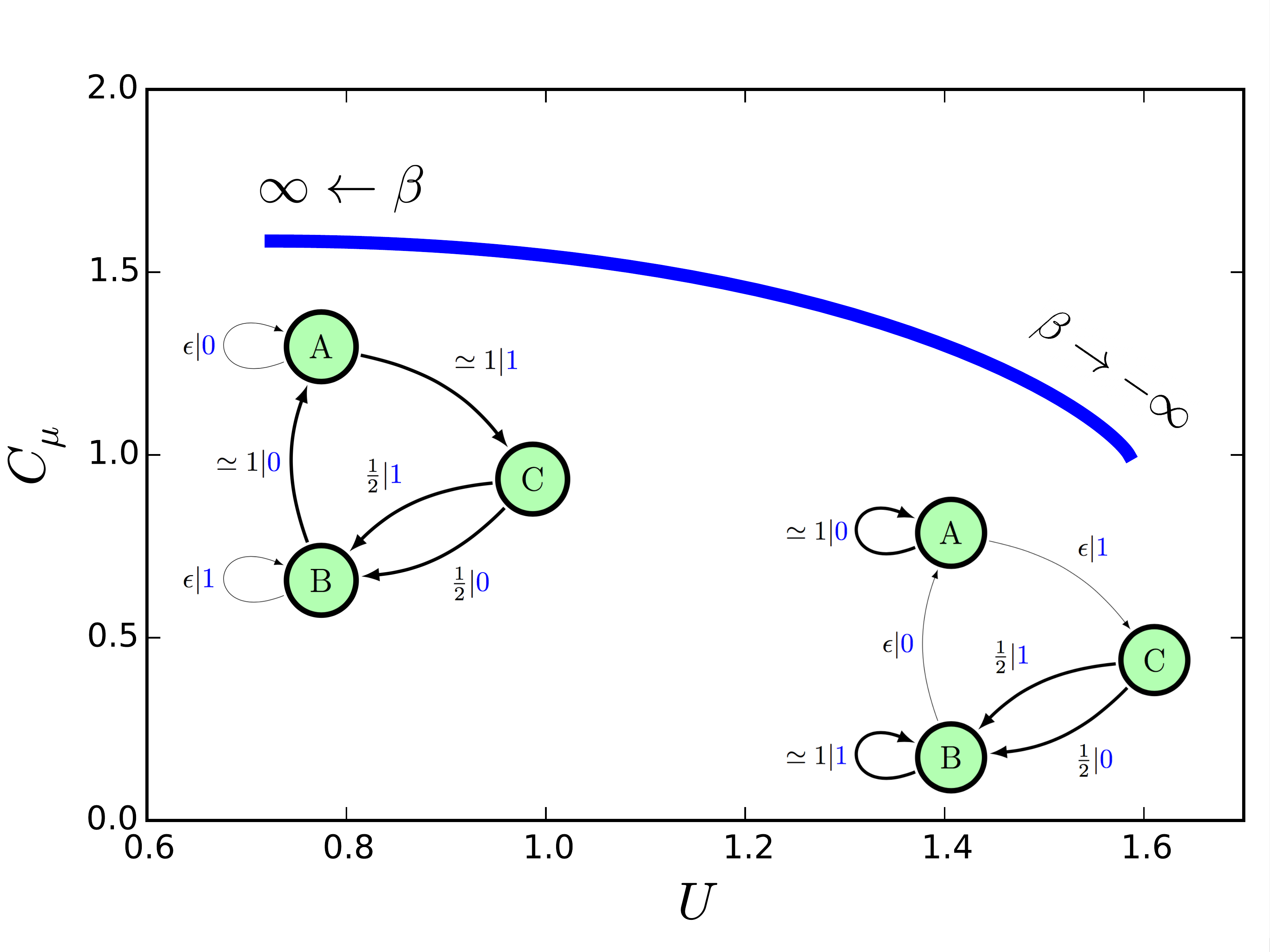}
\caption{(Bottom) Statistical complexity $\Cmu$ versus energy $U$
	for the \EM\ generator (top). The insets (bottom) display the \eMs\ for
	the processes generating the fluctuation extremes at $\beta \to \infty$ and
	$\beta \to -\infty$.
	}
\label{HMM3S}
\end{figure}

Second, this is not the general case, since $\Cmu(\Process^U)$ can be a
nonconstant function of $U$, as we now show. Consider the Intermittent Periodic
Process (IPP) with $p=0.35$; its \eM\ is given in Fig.~\ref{HMMPC}(top right). It gets its
name since when $p=0$, it periodically emits the subsequence $101$ and when $p >
0$, it randomly inserts $1$s. Using the $\beta$-map and Thm.~\ref{THEO} we can
find the processes $\Process^U$ and calculate their $\Cmu$. Figure~\ref{HMMPC}
shows how their $\Cmu(\Process^U)$ depends on $U$. The IPP is similar to the TBC
Process in that it also has a finite energy range; IPP energies $U \in [\approx
0.207,\approx 1.515]$. It turns out that for any process with a finite \eM\ the
allowed energy range is also finite. In addition, the $U$ associated with
$\Process$'s typical set is marked in the figure with a dashed (green) vertical
line near $U \approx 0.406$.

Thus, IPP's $\Cmu(\Process^U)$ is a nontrivial function of $U$. Practically,
this means that generating various rare-sequence classes requires less memory
than for other classes. For example, for events with $U_\text{max}$---$p = 1$
and $\beta \to -\infty$---ones needs no memory, since the class of maximum
energy has only one sequence---the all-$1$s sequence. This can be generated by
an IID process that emits only $1$s. Generally, due to its IID character we do
not need to remember or store the process' current state. In other words, the
\eM\ $M(\Process^U)$ that generates this class only has one state and so $\Cmu
= 0$ bits there. For $U_\text{min}$, occurring at $p = 0$ and $\beta \to
\infty$, there are three ``ground state'' sequences---the three shifts of
$\ldots 101101 \ldots$ and three equally probable states. Thus, $\Cmu
(U_\text{min}) = \log_2 3 \approx 1.585$ bits are necessary for generation.

Third and finally, for a more complex example consider the process generated by
the \eM\ with $p= 1 / 3$ given in Fig.~\ref{HMM3S}(top). Using the $\beta$-map
and Thm.~\ref{THEO} we again find the processes $\Process^U$ and calculate
their $\Cmu$, as shown in Fig.~\ref{HMM3S}(bottom). The difference between this
process and IPP is that at no inverse temperature $\beta$ do we have an IID
process $\Process_\beta$. As a consequence $\Cmu(\Process^U)$ is nonzero for
all allowed $U$.

The insets in Fig.~\ref{HMM3S}(bottom) highlight the details of the process'
\eMs\ for two limits of $\beta$. In the limit $\beta \to \infty$ the
probability of $B$'s self-transition vanishes and the probability of transiting
from state $B$ to $A$ goes to one. Similarly, the probability of $A$'s
self-transition vanishes and the $A$-to-$C$ transition probability goes to one.
As a consequence, as shown in Fig.~\ref{HMM3S}, the extreme process generates
$0$ then $1$, then flips a coin to decide the outcome and then repeats the same
steps again and again.

In the complementary limit $\beta \to -\infty$, an interesting property
emerges. The process breaks into two distinct subprocesses that link to each
other only very weakly. The first process consists of state $B$ with a
deterministic self-transition that generates $1$s. And, the second subprocess
consists of state $A$ with a deterministic self-transition that and generates
$0$s. In other words, the process has two phases that rarely switch between
themselves. As a result, over moderate durations the process exhibits
nonergodic behavior.  We note that this has profound effects on predictability:
substantial resources are required for predicting nonergodic processes
\cite{Crut15a}, despite their requiring finite resources for generation.

\section{Concluding Remarks}

To generate the rare behaviors of a stochastic process one can wait, if one
wants, for exponentially long times for them to occur. Here, we introduced an
alternative to rare-event generation from large deviation theory and its
predecessors. Given a process, we first classified its events into those that
are forbidden, typical, and atypical. And, then we refined the atypical class.
For any chosen rare class we introduced an algorithm that constructs a new
process, and its unifilar HMM, that typically generates those rare events.
Appealing to the optimality of computational mechanics' \eMs\ then allowed us
to analyze the minimal memory costs of implementing rare-event generators.
Depending on the goal---producing a single correct sample (sequential generation) or a
large number of correct of samples (simultaneous generation) from the rare
class of interest---memory cost differs. We studied both costs. Taken together
the three examples analyzed give a complete survey of applying the method and
how memory costs vary across classes of rare events.

The introduction emphasized that we only focused on unifilar HMMs as process
generators and then we constructed the minimal unifilar generator for a given
class of rare events. The unifilar condition is necessary when using a process'
past behavior to optimally predict its future \cite{Maho15a}. However, one may
not be interested in prediction, only generation for which unifilarity is not
required. While removing unifilarity expands the space of HMMs, it greatly
complicates finding minimal generators. For one, \emph{nonunifilar} HMMs can be
more memory efficient than unifilar HMMs for a given process
\cite{Uppe97a,Crut92c,Lohr09b}. For another, constructing a minimal nonunifilar
HMM for a general process is still an open and hard question
\cite{Lohr09c,Lohr12,Gmei11a}.

The required memory $\Cmu(\Process)$ for (unifilarly) generating realizations of
a given process $\Process$ has been used as a measure of structural complexity
for over two decades. It places a total order over stochastic-process space,
ranking processes by the difficulty to generate them. The theorem introduced
here extends the measure $\Cmu(\Process)$ to the full memory spectrum
$\Cmu(\Process^U)$ to generate fluctuations.

As one consequence, this structural accounting introduces the new phenomenon of
the \emph{ambiguity of simplicity} \cite{Agha16a} to the domain of fluctuation
theory. Say that process $A$ is simpler than process $B$, since it requires
less memory to generate: $\Cmu(A) < \Cmu(B)$. However, if instead we are
interested in the rarest events at $U$, we showed that it is possible that $A$
is more complex than process $B$ since it requires more memory for that event
class: $\Cmu(A^U) > \Cmu(B^U)$. As Ref. \cite{Agha16a} notes, this fundamental
ambiguity flies in the face of appeals to simplicity via Occam's Razor and
practically impacts employing statistical model selection as it relies on a
total order of model complexity.

The same fluctuation theory has recently been used to identify fluctuations in
macroscopic thermodynamic functioning in Maxwellian Demons \cite{Crut16aa}.
Moreover, the method can be applied to many stochastic systems to explore their
rare behaviors, from natural processes observed in fluid turbulence
\cite{Anva13a,Tsuj03}, physiology \cite{Kuu04,Prus07}, surface science
\cite{Waec04,Chua05}, meteorological processes \cite{Sura03b}, cosmic microwave
background radiation \cite{Mov11}, seismic time series \cite{Mans09} to
designed systems found in finance \cite{Agha13,Agha14,Ghas07,Huan15}, renewable
energy \cite{Taba14a,Anva16a}, and traffic \cite{Kris02,Naga00}. It gives a
full description of a process,  from its typical to its rare behaviors. And, it
determines how difficult it is to simulate a process' rare events.

Finally, there is another potentially important application domain. The rapid
progress in quantum computation and information suggest that, perhaps soon
even, one will be able to generate processes, both classical and quantum,
using programmable quantum systems. The equivalent memory $C_q$ for the
simultaneous quantum simulation of processes also has already been introduced
\cite{Maho15a,Riec15b,Gu12a,Tan14,Agha16b}. And so, a sequel will analyze
quantum memory fluctuation spectra $C_q(U)$ and how they differ from the
classical spectra introduced here.

\section*{Acknowledgments}

We thank Mehrnaz Anvari and John Mahoney for useful conversations. This
material is based upon work supported by U. S. Army Research Laboratory and the
U.S. Army Research Office under contract W911NF-13-1-0390.

\appendix 
\section{Proof of the Theorem}
\label{app:ThmProof}

This appendix establishes the main theorem via a single lemma relying on
a process' cryptic order.

\emph{Cryptic order} is a recently introduced topological property of
stochastic processes \cite{Crut08a} that is bounded by, but is rather different
in motivation from, the more familiar Markov order \cite{Mehra89}. Formally,
given a process' \eM, its \emph{cryptic order} is $K=\inf \big\{l: \H[\St_l |
X_0X_1\cdots] = 0 ,\ l \in \mathbb{Z}\big\}$. Informally, this means that if we
observe an infinite length realization, we can be certain about in which state
the \eM\ is in after the $K^{th}$ symbol \cite{Crut08b}.

{\Lem For any given process with finite states and cryptic order, for every $U$
and $\beta \in \mathbb{R}/0$ we have:
\begin{align*}
\Lambda^\Process_U  = \Lambda^{\Process_\beta}_{\beta U - \log_2 \MaxEigBeta}
  ~.
\end{align*}
}

\begin{figure}
\centering
\vspace{10 mm}
\includegraphics[width=1\columnwidth]{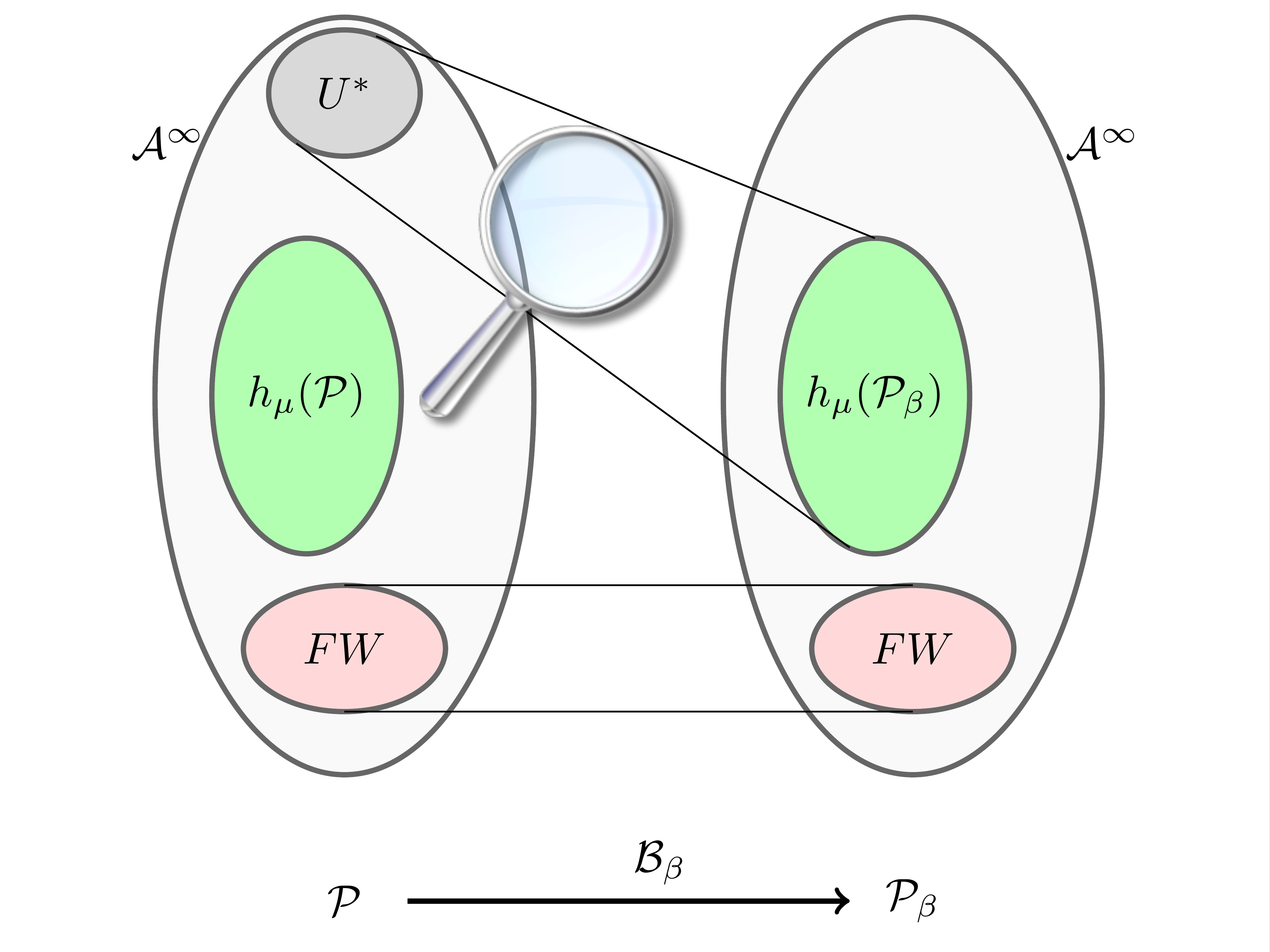}
\caption{The $\beta$-map acts like a magnifier: In the parlance of large
	deviation theory, it ``twists'' or ``tilts'' the sequence distribution in a
	way that focuses on the probability of a chosen rare-event class. Fixing
	$\beta$, the $\beta$-map changes the energy $U$ of a class to $U_\beta =
	\beta U - \log_2 \MaxEigBeta$. In particular, a subset with energy $U^*$
	maps to the typical set of a new process that has energy
	$\hmu(\Process_\beta)$. The set FW of forbidden sequences is invariant under
	the $\beta$-map.
	}  
\label{BMAP}
\end{figure}

{\ProThe
Consider an arbitrary word $w=\ms_0 \ms_1 \ldots \ms_{n-1} \in \MeasAlphabet^n$
generated by process $\Process$ where $n \gg 1$.  
Since the \eM\ is unifilar, immediately after choosing the initial state, all the successor
states are uniquely determined. Using this, we can decompose $w$ to two parts:
The first part $w_K$ is the first $K$ symbols and the second part is $w$'s
remainder. Knowing $w$, the state $\st_K$ and all successor states following
$\st_{K+1},\st_{K+2},\ldots$ are uniquely determined. As a consequence, the
probability of process $\Process$ generating $w$ can be written as:
\begin{align*}		
\mathbb{P}(w)= \mathbb{P}(w_K) \prod_{i=K}^{n-1}
  \left( {\textbf T}^{(\ms_i)} \right)_{\st_i \st_{i+1}}
  ~.
\end{align*}
We can adapt the energy definition in Eq.~(\ref{ENERGY}) to finite-length
sequences. Then, $w$'s energy is:
\begin{align*}		
\mathcal{U}(w) & = - \frac{\log_2 \mathbb{P}(w)}{n}\\
     & = - \frac{\log_2 \mathbb{P}(w_K)}{n}
  -\frac{\log_2 \left( \prod_{i=K}^{n-1} \left({\textbf T}^{(\ms_i)}
  \right)_{\st_i \st_{i+1}} \right)}{n}
   ~.
\end{align*}
Now consider the same word, but this time generated by the \eM\
$M(\Process_\beta)$. Then, the probability of generating $w$ is:
\begin{align*}		
\mathbb{P}_\beta(w)
  & = \mathbb{P}_\beta(w_K)
  \prod_{i=K}^{n-1}
  \left({\textbf S}_\beta^{(\ms_i)} \right)_{\st_i \st_{i+1}} \\	
  & =\mathbb{P}_\beta(w_K)
  \prod_{i=K}^{n-1}
  \frac{\left({\textbf T}_\beta^{(\ms_i)}\right)_{\st_i \st_{i+1}} 
  (\MaxRvecBeta)_{\st_{i+1}}}{\MaxEigBeta (\MaxRvecBeta)_{\st_i } } \\
  & = \mathbb{P}_\beta(w_K)
  \frac{(\MaxRvecBeta)_{\st_{n}}}{(\MaxRvecBeta)_{\st_K}}
  \left(\MaxEigBeta \right)^{n-K}
  \prod_{i=K}^{n-1}
  \left({\textbf T}_\beta^{(\ms_i)}\right)_{\st_i \st_{i+1}} \\
  & = \mathbb{P}_\beta(w_K)
  \frac{(\MaxRvecBeta)_{\st_{n}}}{(\MaxRvecBeta)_{\st_K}}
  \left(\MaxEigBeta \right)^{n-K}
  \left( \prod_{i=K}^{n-1} \left({\textbf T}^{(\ms_i)}\right)_{\st_i \st_{i+1}}
  \right)^\beta
  ~.
\end{align*}
The new energy for the same word is:
\begin{align*}		
\mathcal{U}_\beta(w)= &- \frac{\log_2 \mathbb{P}(w)}{n} \\
  & = - \frac{\log_2 \left(\mathbb{P}_\beta(w_K)\frac{(\MaxRvecBeta)_{\st_{n}}}
  {(\MaxRvecBeta)_{\st_K}} \right)}{n}
  - \frac{n-K}{n}\log_2 \MaxEigBeta \\
  & \qquad \qquad - \beta
  \frac{\log_2 \left(\prod_{i=K}^{n-1} \left({\textbf T}^{(\ms_i)}\right)_{\st_i \st_{i+1}}\right)}{n}
  ~.
\end{align*}
In the limit of large $n$ the first terms in $\mathcal{U}(w)$ and
$\mathcal{U}_\beta(w)$ vanish and we have $\mathcal{U}_\beta(w) = \beta
\mathcal{U}(w) - \log_2 \MaxEigBeta$. Thus, for any two long sequences $w_1, w_2
\in \MeasAlphabet^n$, if $\mathcal{U}(w_1)=\mathcal{U}(w_2)$, then
$\mathcal{U}_\beta(w_1) = \mathcal{U}_\beta(w_2)$. And, the partitions induced
by Eq.~(\ref{ENERGY}) are invariant under the $\beta$-map. In other words,
the energy of an arbitrary bubble after $\beta$-mapping changes from $U$ to
$U_\beta$, where:
\begin{align*}
U_\beta = \beta U - \log_2 \MaxEigBeta
  ~.
\end{align*}
This completes the lemma's proof.
}

This demonstrates how the $\beta$-map changes bubble energy: $U \to \beta U -
\log_2 \MaxEigBeta$. So, now we ask for the bubble (and its energy) that maps
to the typical set of the new process $\Process_\beta$. That is, we use the
$\beta$-map to find the class $\Lambda^\Process_U$ of rare sequences typically
generated by $M(\Process_\beta)$.

This sets up the theorem's proof. Using the fact that the process' metric
entropy is the typical set's energy, the energy of $\Process_\beta$'s typical
set is $\hmu(\Process_\beta)$. (Refer to Fig.~\ref{BMAP}.) The lemma tells us
how the $\beta$-map changes energy. Using this, we can identify the bubble with
energy $U^*$ that is typically generated by $M(\Process_\beta)$, it has:
\begin{align*}		
\hmu(\Process_\beta) = \beta U^* - \log_2 \MaxEigBeta
  ~.
\end{align*}
This completes the theorem's proof.

\bibliography{chaos}

\end{document}